\documentstyle[12pt,aps,prb]{revtex}

\def\i{{\rm i}}
\def\e{{\rm e}}
\def\h2o{H$_2$O}
\def\d2o{D$_2$O}

\def\etal{{\em et al.}}

\def\mathcal{\cal}

\begin{document}

\title{Decoherent Histories and Non-adiabatic Quantum 
 Molecular Dynamics
Simulations.}
\author{Eric R. Bittner}
\address{Department of Chemistry, University
of Houston, \\Houston TX 77204}
\author{Peter J. Rossky}
\address{Department of Chemistry and Biochemistry,
University of Texas at Austin,\\ Austin, TX 78712\\
 }

\date{\today}
\maketitle

\begin{abstract}
The role of quantum coherence loss in mixed quantum-classical
dynamical systems is explored in the context of the theory of quantum
decoherence introduced recently by Bittner and Rossky.  (J. Chem.
Phys. {\bf 103}, 8130 (1995)).  This theory, which is based upon the
consistent histories interpretation of quantum mechanics, introduces
decoherence in the quantum subsystem by carefully considering the
relevant time and length scales over which one must consider the
effects of phase interference between alternative histories of the
classical subsystem.  Such alternative histories are an integral part
of any quantum-classical computational scheme which employ transitions
between discrete quantum states; consequently, the coherences between
alternative histories have a profound effect on the transition
probability between quantum states.  In this paper, we review the
Bittner-Rossky theory and detail a computational algorithm suitable
for large-scale quantum molecular dynamics simulations which
implements this theory.  Application of the algorithm towards the
relaxation of a photoexcited aqueous electron compare well to previous
estimates of the excited state survival time as well as to the
experimental measurements.
\end{abstract}

\section{Introduction}

A key issue which has emerged in describing electronic transitions in 
the condensed phase is the proper treatment of the short-lived phase 
coherence of the electronic and nuclear (bath) wavefunction.  The 
essence of the effect becomes clear if one considers the behavior of 
an electronic quantum system in a classical bath.  In such a mixed 
quantum-classical system, the classical dynamics of the bath nuclei 
follow along a given adiabatic potential surface associated with one 
of the eigenstates of the electronic Hamiltonian.  Thus the potential 
felt by the nuclear degrees of freedom will depend strongly upon the 
quantum state of the electronic degrees of freedom.  The fundamental 
distinction between quantum mechanics and classical mechanics is that 
a quantum system can evolve into a coherent linear superposition of 
states.  This leads to a rather profound ambiguity in describing the 
forces between the quantum and classical subsystems.  Furthermore, 
quantum mechanics forces us to consider the effect of all alternative 
histories or pathways, including those of the classical bath, when 
computing quantum transition probabilities.  Thus, as the electronic 
wavefunction evolves from an initially pure eigenstate to a coherent 
superposition of eigenstates, various alternative pathways for the 
bath particles begin to emerge, each associated with dynamics starting 
from an initial nuclear configuration and electronic state and ending 
in a different final nuclear configuration and final electronic 
eigenstate.  As the nuclear dynamics for the different 
quantum (electronic) states diverge, phase coherence between the 
alternative histories is lost due to the increasing differences 
between the classical actions associated with each alternative 
nuclear path.  This effect is known as quantum 
decoherence.~\cite{OmnesIQM}

Because quantum decoherence acts to suppress the formation of 
superposition states, it diminishes profoundly the transition 
probability between quantum states coupled by the nuclear dynamics.  
When the loss of quantum phase coherence between the electronic states 
is neglected and no longer enters directly into the computation of 
making an electronic transition, which is a typical approximation made 
in mixed quantum-classical treatments, one can show that the result 
will be an incorrect estimation of the transition probabilities and 
other associated physical observables.\cite{ERB95b}  Because of the 
tremendous utility of mixed-quantum classical treatments in simulating 
condensed phase phenomena, it is of tantamount importance to be able 
to properly incorporate the effects of quantum decoherence into these 
simulations.

Recently, we have presented a theory of quantum decoherence which is 
suitable for condensed phase quantum molecular dynamics simulations.
\cite{ERB95b}  
This approach is based upon a interpretation of quantum 
mechanics introduced by Griffiths~\cite{Griffiths84}, 
Omn{\`e}s~\cite{Omnes88,Omnes89}, and Gell-Mann and 
Hartle~\cite{MGM90,MGM93,MGM94} over roughly the past 10 years.  This 
interpretation of quantum mechanics, termed ``consistent'' or 
``decoherent'' histories, has generated a great deal of attention in 
the field of quantum 
cosmology.\cite{Deco1,Deco2,Halliwell95a}  We have
also applied and extended the formalism to study coherence effects in 
physically realistic condensed phase problems ~\cite{ERB96a,ERB96b}.  
This work has lead to a number of advances in designing computational 
algorithms which incorporate a consistent description of quantum 
decoherence and to a deeper understanding of the role that quantum 
decoherence plays in condensed phase chemical processes and molecular 
level descriptions of the decoherence 
process.~\cite{ERB95b,ERB96a,ERB96b,PJR97a}  These theoretical advances have 
also offered a tangible solution to a specific long standing puzzle 
regarding the solvent isotope effect on the non-radiative lifetime of 
an excess electron in \h2o and \d2o and offer a plausible 
explanation of the lack of an observed solvent isotope 
effect.\cite{Gaud87,Barb93a,Barb94a,Barbara93}

In this paper, we give the details of our computational 
algorithm based upon consistent quantum histories and its application 
in a full scale quantum molecular dynamics simulation of the 
relaxation of an excited excess electron in \h2o.  The simulations 
presented in this paper represent the first application of the 
Bittner-Rossky decoherence theory in a full scale simulation of a 
condensed phase system.  The aqueous electron provides perhaps the 
simplest example of electronic dynamics in the condensed phase.  At 
the same time, aqueous electrons are of particular importance since 
they play a prominent role in the broad area of radiation chemistry of 
water.  In spite of intensive experimental and theoretical study since 
identification of the hydrated electron over 30 years ago, many 
dynamical features of electron solvation have remained incompletely 
understood, largely due to the extremely fast time scales on which 
electronic relaxation occurs.\cite{Gaud87}  Newly developed 
femtosecond spectroscopic techniques are now providing glimpses of the 
details of condensed phase electronic dynamics.  From a theoretical 
standpoint, the aqueous electron is an ideal test case for new 
condensed phase theories.  Recent theoretical advances in treating 
condensed phase electronic 
dynamics~\cite{PJR90,PJR91a,PJR91b,BJS94c,BJS94d} has led to an 
exceptional interplay between theory and experiment, helping to 
unravel many subtle dynamical aspects of this system.

The remainder of this paper proceeds as follows: In the next section 
we review the consistent histories methodology as it applies toward 
developing a computational algorithm for condensed phase simulations.  
We then give the results of quantum MD simulations of the 
non-adiabatic relaxation dynamics of an excited excess aqueous 
electron which employ the consistent histories algorithm and compare 
the present results to previous estimates of the non-adiabatic 
electronic transition rate by Schwartz, Bittner, Prezhdo, and 
Rossky (SBPR)~\cite{ERB96a}, to earlier results presented by Schwartz and 
Rossky which employed a constant 1 fs coherence time scale 
throughout,\cite{BJS94c,BJS94d,BJS94e} as well as to the experimental
data of  Kimura~\etal. \cite{Barb93a,Barb94a,Barbara93} 
Finally, we comment upon the utility of 
the present algorithm and discuss its limitations as well as future 
extensions and improvements.

\section{Theoretical Methods}

\subsection{Coherence Between Switching Paths}

Let us consider the evolution of an initial quantum state $| i 
(R_\circ)\rangle$, taken as an adiabatic eigenstate of the quantum 
mechanical Hamiltonian for initial bath (or classical) configuration 
$R_\circ$, along a switching path, $R^\alpha(t)$ which begins at 
$R_{\circ}$.  Along the course of this path, we determine at various 
time intervals whether or not a quantum transition has occurred 
according to a stochastic switching criteria and modifying the 
classical dynamics accordingly.  
The fundimental notion of having classical particles switch between
discrete quantum states was originally pioneered by
Tully~\cite{Tully71} in the early 1970's.  Subsequently, various
computational schemes have been developed which incorporate quantum
switching into the classical dynamics and improve upon Tully's
original idea.  Most notably are the stationary phase surface hopping
algorithms developed by Webster, Friesner, and
Rossky~\cite{PJR91a,PJR91b} and further by Coker and
co-workers~\cite{Coker93,Coker94,Coker91,Coker92}, as well as the
Molecular Dynamics with Quantum Transitions algorithm of Tully and
co-workers~\cite{Tully90,Tully94}. 
The classical switching path contains 
a record of the outcome of each switching attempt and can thus be 
written as a time ordered sequence of events,
\begin{eqnarray}
R^\alpha(t) = \{R_\circ^{\alpha_\circ},
\cdots, R_j^{\alpha_j} \cdots, R_n^{\alpha_n}\}.\label{eq:path}
\end{eqnarray}
The superscript $\alpha_j$ denotes the switching outcome at time-step
$j$ and $\alpha_\circ = i$ corresponding to our choice of the initial
quantum state. Two such paths are shown schematically in
Fig.~\ref{fig:2paths} where we plot the eigenenergy of the occupied state
along the path as a function of time.  Changes in the quantum state
imply that there is a corresponding sudden change in the forces
exerted on the classical particles over the course of its evolution.
The result is that different sequences of switching events will lead
to rapidly diverging paths.

Along a given switching path, $R^\alpha(t)$, the partial 
transition amplitude between an initial and final quantum state,
$|i(R_\circ)\rangle$ and $|f(R_f)\rangle$ respectively, is given by
\begin{eqnarray}
T_{if}[R^\alpha(t)] &=& \langle f (R^\alpha_f) |
\e^{-\i\int_{t_\circ}^{t_f} ds H_Q[R^\alpha(s)]}| i (R_\circ) \rangle
\e^{+\i S[R^\alpha(t)]}\nonumber \\ 
&=& U_{if}[R^\alpha(t)]\e^{+\i S[R^\alpha(t)]} ,~\label{tpart}
\end{eqnarray}
where $U_{if}[R^\alpha(t)]$ is the transition amplitude for the
quantum subsystem (with Hamiltonian $H_{Q}$)  
and $S[R^\alpha(t)]$ is the classical action
computed along the switching path $R^\alpha(t)$, i.e.
\begin{eqnarray}
S[R(t)] = \int_{t_\circ}^{t_f} dt\left\{ \frac{m}{2}\dot{R}(t)^2 -
 V_C(R(t))\right\}.
\end{eqnarray}
Here, $V_{C}$ is the interaction potential coupling the classical 
particles. 
The probability of starting in some initial state
$|i(t_\circ)\rangle $ ending up in a given final quantum state, $|
f(t_f)\rangle $, at some time $t_f$ is computed by summing over all
possible switching paths which connect the initial and final quantum
states. Averaging over initial
configurations, the reduced transition probability for a mixed
quantum-classical system is
\begin{eqnarray}
P_{if}(t_f)& = & \left\langle \sum_{\{R^\alpha(t),R^{\beta}(t)\} }
T_{if}[R^\alpha(t)] T^\dagger_{if}[R^\beta(t)] \rho_i(R_\circ)
\right\rangle_{\circ}, \nonumber \\ &=& \left\langle
\sum_{\{R^\alpha(t),R^{\beta}(t)\} }
U_{if}[R^\alpha(t)]\rho_i(R_\circ))U^\dagger_{if}[R^\beta(t)] \e^{\i(
S[R^\alpha] - S[R^\beta])}\right\rangle_{\circ},
\label{eq:prob}
\end{eqnarray} 
where $\rho_i(R_\circ)$ is the probability density of being in the
initial state, the average, $\langle \cdots\rangle_\circ$,
is taken over initial configurations, and
the sum is over pairs of switching paths
\begin{eqnarray}
\{R^\alpha(t), R^\beta(t) | R^\alpha(0) = R_\circ, R^\beta(0) =
R_\circ\}
\end{eqnarray}
which start at the initial configuration, $R_\circ$, with the quantum
state in the initial state and end at any final configuration with the
quantum state in state $|f\rangle$ at $t_f$.   

According to this last equation, the transition probability for the 
quantum subsystem is dependent upon interferences between the various 
alternative pairs of switching pathways.  Thus, in order to correctly 
compute the populations of the electronic states at some later time, 
one must perform a sum over all alternative paths that the bath can 
take.  In a practical sense, this amounts to launching a swarm of 
trajectories from each initial starting point and computing the 
switching probabilities (i.e. branching ratios for the swarm) 
using Eq.~\ref{eq:prob}.
However, this approach has the distinct disadvantage that one must 
initialize a large enough swarm of trajectories in order to 
effectively sample the distribution of switching paths originating 
from each initial configuration.  So, while formally correct and 
tractable in small dimensional systems, the tremendous computational 
overhead required to implement Eq.~\ref{eq:prob} directly for a large 
number of classical variables is prohibitive. 

Next, we present an alternative derivation  of the quantum survival 
and transition probabilities which uses only a single switching path
explicitly.  The formalism is motivated by the observation that  
at short times, there will be 
a significant contribution from the phase interferences between two 
alternative paths with similar histories, necessitating a 
coherent sum over paths, while at longer times the action difference 
between two paths will become very large and phase interference 
effects from alternative paths 
will become negligible.  The time scale separating these two 
regimes is the quantum coherence time scale. 

\subsection{Consistent Histories}
Let us define the space of paths, ${\mathcal Z}$, as consisting of all 
continuous paths connecting some initial system/bath state 
$R_\circ^{\alpha}$ to the final state $R_f^{\beta}$ labeled by a set 
of quantum numbers, $\{\alpha_j\}$ and a classical bath variable, $R$.  
Define also a set of subspaces of ${\mathcal Z}$ 
labeled by $\Delta(R^{\alpha})$ which are the spaces of paths taken by the
quantum  subsystem which are parameterized by the paths taken by the bath.  
The full space of paths is the continuous union of all the 
various subspaces.
\begin{equation}
	{\mathcal Z} =\biguplus_{\alpha} \Delta(R^{\alpha})
\end{equation}
These paths are {\em completely fine grained histories} since the 
values of the each paths can be specified at all times and one can 
compute the partial amplitude along any one of the paths 
in ${\mathcal Z}$.  Furthermore, by integrating over the individual 
``quantum''  subspaces
one obtains Eq.~\ref{tpart},
\begin{eqnarray}
  	T_{\alpha\beta}[R^{\alpha}] & = & \int_{\Delta(R^{\alpha})}Dx
\exp\left(\i (S_{Q}[x,R^{\alpha}] + S[R^{\alpha}]\right) \nonumber \\
   &=&  U_{if}[R^\alpha(t)]\e^{+\i S[R^\alpha(t)]}.
\end{eqnarray}

A fundamental property of quantum mechanics is that fine grained 
histories can not be assigned probabilities, only amplitudes.  In 
order to make predictions based upon the theory of probabilities, {\em 
coarse graining} of histories must be introduced.  Coarse graining is 
accomplished when a ``reduced'' history is constructed by summing over 
sets of fine grained histories at a particular time.  Coarse grained 
histories can be assigned probabilities, that is, integrating over 
coarse grained histories is equivalent to integrating over 
probabilities. 

Coarse graining of histories arises naturally whenever we try to mix 
quantum with classical dynamics.  As different classical 
switching paths emerge 
from a common origin, the space of quantum paths parameterized by one 
switching path will diverge from the space of quantum paths 
parameterized by another.
As the ``overlap'' between subspaces 
decays (corresponding to different classical dynamics),
the originally fine grained space of paths, ${\mathcal Z}$, becomes  
coarse grained into separate subspaces, $\{\Delta(R^{\alpha})\}$
as time goes on.  Consequently,
transition probabilities are computed by first 
integrating coherently  over 
the fine grained sets of paths  to get the transition
amplitude say along one classical switching path, then summing 
incoherently (i.e. adding as ordinary probabilities) over coarse 
grained sets.

Quantum decoherence effects 
can be consistently incorporated into mixed quantum-classical systems 
by recognizing that restricting the quantum evolution to given 
classical pathways is equivalent to making a series of quantum 
measurements on the total system.  Furthermore, the quantum coherence time 
is the time scale which characterizes how often such ``measurements'' 
occur~\cite{ERB95b}.  The classical path sequence shown 
in Fig.~\ref{fig:2paths} is an example of a quantum mechanical 
history.  Along this history, the time evolution operator for the 
quantum system can be written equivalently as a time ordered sequence 
of alternating quantum projection operators and unitary evolution 
operators
\begin{eqnarray}
\hat C[R^\alpha(t)]  =
\hat U_n \hat P^{\alpha_{n-1}} \cdots \hat U_1 P^{\alpha_1} \hat U_\circ
\end{eqnarray}
where the projections  at each time interval are members of complete
sets representing the total set of possible outcomes,
\begin{eqnarray} 
\sum_{\alpha_i} \hat P^{\alpha_i} = 1,
\end{eqnarray}
 and
\begin{eqnarray}
\hat U_n = \e^{-\i \int_{t_{n-1}}^{t_n}ds H[R^{\alpha}(s)]}
\end{eqnarray}
is the unitary evolution operator for the quantum wavefunction along a
segment of the switching path.

The decoherence 
time scale sets a characteristic time interval between subsequent 
applications of the projection operators.  This time scale is roughly 
the time scale for alternative stationary trajectories of the nuclear 
bath to diverge sufficiently such that the difference in action 
between the two paths in Eq.~\ref{eq:prob} is large.  
In Ref.12 we suggested that coherence time scale is related to the 
time scale for the decay of the overlap integral between nuclear 
wavefunctions evolving on different adiabatic potential energy 
surfaces. During the coherence interval,
transition amplitudes are added coherently according to the 
rules of quantum mechanics.  Application of the projection operators 
destroys the coherences and one is  left with transition probabilities 
which are then added according to the rules of standard probability 
theory.  Within this approach, Eq.~\ref{eq:prob} is equivalent to 
\begin{eqnarray}
P_{i\rightarrow f}(t) &=& \left\langle
\sum_{\{R^\alpha(t)\}} \langle f_n|
\hat C[R^\alpha(t)] \rho_i(R_\circ) 
\hat C^\dagger[R^\alpha(t)] | f_n\rangle
\right\rangle_{\circ} \nonumber \\
& =  & 
\left\langle 
\sum_{\{R^\alpha(t)\}}
\langle f_n | 
\hat U_n
\hat P^{\alpha_{n-1}}\cdots
\hat P^{\alpha_1}
\hat U_1
\rho_i^\circ
\hat U^\dagger_1
\hat P^{\alpha_1}\cdots
\hat P^{\alpha_{n-1}}
\hat U^\dagger_n
| f_n \rangle
\right\rangle_{\circ},\label{eq:ch-prob}
\end{eqnarray} 
where $|f_n \rangle = |f(R^{\alpha_n})\rangle$ is the final quantum 
state at the end of path $R^{\alpha}(t)$.

When the time intervals $\delta t_n = t_n - t_{n-1}$ are members of a 
Poisson distribution, the probability of maintaining coherence over a 
given interval, $t$, follows from the exponential deviate
\begin{eqnarray}
\pi(t) = \int_0^{ t} ds \frac{\e^{-s/\tau}}{\tau}
= 1 - \e^{-t/\tau}, \label{eq:poisson}
\end{eqnarray}
and $\tau$ is the characteristic decoherence timescale.  
Thus, over a short time 
interval, $\delta t$, where the probability of collapsing the 
wavefunction is
\begin{eqnarray}
	\pi(t) & = & \delta t/\tau  +{\mathcal O}(\delta t^{2}),
 	\label{prob}
\end{eqnarray}
the reduced density matrix of the quantum system evolves as
	\begin{eqnarray}
\rho(t+\delta t) = ( 1 - \delta t/\tau)(\rho(t)+i\delta t [H,\rho])
+ \delta t/\tau{\mathcal R}[\rho].~\label{finite}
\end{eqnarray}
where $\mathcal R[\rho]$ is a ``reduction mapping'' of the density 
matrix $\rho$ which projects out the diagonal elements of the 
quantum density matrix (the populations) killing off the coherences 
between the quantum states.

Taking the limit of $\delta t\rightarrow 0$ 
in Eq.~\ref{finite} and identifying $\tau$ as the decoherence 
timscale $\tau_{D}$
we arrive at the  master equation for the quantum density matrix.
\begin{eqnarray}
\dot{\rho}(R)&=& \i[H(R),\rho(R)] 
 - \frac{1}{\tau_{D}} (\rho - {\mathcal R}[\rho(R)]),\nonumber \\
&=&{\mathcal L}(R)\rho - \frac{1}{\tau_{D}} (\rho(R) - {\mathcal R}[\rho(R)]).
\label{eqn.lvn}
\end{eqnarray}
The first term in this equation contains the Liouvillian of the 
quantum system, ${\mathcal L}(R) =\i[H(R),\,]$.  Evolution of the 
quantum system under this term alone is unitary and non-dissipative.  
The second term introduces {\em quantum decoherence} into the dynamics 
of the quantum subsystem.  The coherences originally in the quantum 
subsystem decay due to the series of measurements imposed by the 
environment.  Although energy exchange does not appear explicitly in 
Eq.~\ref{eqn.lvn}, energy exchange between the system and the bath is 
explicitly included through dynamics of the bath variable $R(t)$ 
treated separately.

In other theories of quantum relaxation, such as the spin-boson model 
~\cite{Leggett81,Leggett87}, the Redfield equations 
~\cite{Redfield,Abragam,Jean92,Jean94,Jean95}, or Liouville space 
methods~\cite{MukamelNOS}, both decoherence and dissipation are 
treated implicitly through effective equations of motion for the bath, 
thus losing the molecular level information about the underlying 
dynamics of the bath.  In our treatment, dissipation is included {\em 
explicitly} through the classical molecular dynamics of the condensed 
phase medium, thus we are able to retain a molecular level description 
of the bath while quantum decoherence is treated implicitly in order 
to avoid summing over alternative pairs of classical paths.

It is important to note that the form of the projections and the 
coherence time scale (and the resulting master equation) are directly 
related to the forces coupling the quantum system to the bath and 
hence are very dependent upon the choice of basis used to represent 
the quantum subsystem.  In the present application of our theory, we 
assume that the projections periodically resolve the quantum subsystem 
into the adiabatic states with a time scale determined by the average 
decay time of the Franck-Condon overlap between nuclear wavefunctions 
initialized on the different adiabatic surfaces at the instantaneous 
nuclear coordinates.\cite{ERB96a}  A more detailed analysis of this 
assumption is forthcoming.\cite{PJR97a}

Perhaps the most important physical consequence of decoherence for 
chemical physical applications is the overall reduction of the 
transition probabilities between different quantum states.  In order 
for a quantum system to make a transition from one state to another, 
the system must form a coherent superposition between the states.  The 
ease with which these coherences form reflects the strength of 
the coupling between the two states.  Decoherence, on the other hand, 
reflects the strength of the system-bath coupling. 
When a quantum system evolves into a 
superposition state, the destructive interferences between divergent 
alternative 
switching paths leading to the final state destroys the coherence in 
the quantum subsystem, hence, reducing the likelihood of making a 
transition into the final state.~\cite{ERB95b}

\section{Simulations}

The simulation techniques employed here are nearly identical to those 
used in earlier work by Murphrey and Rossky studying both the 
relaxation of electrons photo-injected into neat 
water~\cite{PJR91a,PJR93} as well as the present case of 
photoexcitation of equilibrium hydrated electrons by Schwartz and 
Rossky. The exceptions regard the treatment of the quantum 
coherences detailed below.~\cite{BJS94d,BJS94e,BJS95a} Briefly, the 
model consists of 200 classical SPC water molecules with the addition 
of internal flexibility~\cite{Rahman85} and one quantum electron in a 
cubic cell of side 18.17 \AA\ (corresponding to a solvent density of 
0.997 g/ml) with standard periodic boundary conditions at room 
temperature.  The electron-water interactions were described with a 
pseudo-potential,~\cite{PJR87} and the equations of motion integrated 
using the Verlet algorithm with a dynamical time step of 1 fs in the 
microcanonical ensemble.~\cite{AT} The adiabatic eigenstates at each 
time step were computed via an efficient iterative and block Lanczos 
scheme utilizing a 16$^3$ plane wave basis~\cite{PJR91a} The lowest 6 
eigenstates were used as a basis for propagating the quantum 
wavefunction and to compute the stationary phase paths for the nuclear 
dynamics.  Explicit details of these methods are discussed 
elsewhere.~\cite{PJR91a}

\subsection{Consistent Histories Algorithm}
Quantum decoherence was incorporated into the simulation through 
the Consistent Histories theory discussed above.  The computational 
algorithm given below follows directly from Eq.~\ref{eq:ch-prob} and 
the coherence time intervals were chosen from a Poisson distribution 
with characteristic time scale $\tau_{D} = 3.1$ fs which was derived 
from our previous estimates of the coherence time scale for an electron 
in \h2o when the excited state is nearly solvated.  As discussed in a 
previous paper, this time scale was estimated by computing the average 
decay time of the overlap of frozen Gaussian vibrational wavefunctions 
evolving on different adiabatic potential energy surfaces by sampling 
a large number of nearly equivalent excited state configurations.  
This is in marked contrast to previous simulations on this system in 
which coherences between electronic states were not maintained beyond 
1 fs.  The consistent histories algorithm proceeds as follows:

\begin{enumerate}
	\item Determine new coherence interval from Poisson distribution 
	of possible intervals with characteristic time scale $\tau_{D}$.  
	\item Propagate the quantum wavefunction over this time scale while 
	self consistently evolving the classical degrees of freedom.  At 
	the end of each dynamical time step $h \le \tau_{D}$, determine the 
	switching path followed by the classical variables using the 
	stationary phase algorithm developed by Webster, {\em et 
	al.}~\cite{PJR91a,PJR91b} modified such that coherence in the quantum 
	wavefunction is maintained over the entire coherence interval.  We 
	also note that the stationary phase switching path is determined 
	in a piece-wise continuous fashion by selecting intermediate 
	quantum states every dynamical time step.  This is to avoid the 
	computational overhead of computing variationally the 
	stationary phase trajectory over relatively long time intervals.  
	So long as $\tau_{D}$ is no longer than a few dynamical time steps, 
	this approximation should not be too extreme; however, certain 
	pathological cases can be invented in which this approximation 
	does break down.
        
	\item If either a switch occurs in the time interval or we reach 
	the end of the interval, the quantum wavefunction is collapsed 
	using the projection operators discussed above.
	
	\item  Repeat.

\end{enumerate}

\subsection{Results}
In Fig.~\ref{fig:swtimes} we plot the switching times from the excited 
state to the ground state for a total of 23 simulation runs.  The 
starting configurations were generated by performing a 35 ps 
simulation in which the electron was prepared in the ground state and 
16 initial configurations were chosen whenever the energy difference 
between the ground state and one of the p-like excited states become 
resonant with the excitation laser (1.7 eV). 
~\cite{Barb93a,Barb94a,Barbara93}
These starting 
configurations were identical to configurations used previously by 
Schwartz and Rossky in their work on this system.
~\cite{BJS94d,BJS94e,BJS95a} Following the 
initial excitation, the system was allowed to evolve.  During this 
time, the energy gap between the p-states and the s-state narrowed to 
$\approx 0.5 eV$ as the excited state was solvated by the surrounding 
water molecules.  The simulation continued until a switch from the 
excited state to the ground state was recorded.  Immediately after the 
switch, the energy gap between the occupied ground state and the first 
excited state widened dramatically as the solvent responded to the new 
electronic state.  The average and median switching times from these 
simulations are compiled in Table 1.

In order to test the sensitivity of the switching times to the choice 
of sequence of coherence intervals, 
five configurations were ``recycled'' by using different 
random number sequences for the coherence time intervals to generate 
different switching paths starting from the same initial configuration.
 For each such pairs of paths, the 
classical dynamics were identical up until the earlier switching time 
when one of the paths switched from the excited state to the ground 
state.  However, since the coherences between the different quantum 
states were killed off at randomly different times for each path, each 
path sampled a different probability distribution function for making 
a switch at each MD time step.  Hence any correlations between 
switching times resulting using the same initial configuration reflect 
correlations in the distribution of coherence time intervals.  Given 
that there is typically a 100 fs time difference between pairs of data 
(in one case a 530 fs difference) which is roughly 1/3 of the average 
survival time scale of the excited state, we find very little 
correlation between the switching times originating from the same 
initial configuration.

The excited state survival probability reflects not only the 
probability of making a transition from the excited state to the 
ground state, it also reflects the solvation dynamics of the 
non-equilibrium excited state following its preparation.  Initially, 
the energy gap between the excited and ground state is very wide and 
a non-adiabatic transition between the states is highly unlikely.  
However at later times, after the excited state has begun to be 
solvated by the surrounding water molecules, the energy gap closes 
considerably and nonadiabatic transitions become more likely.  The 
solvation dynamics of this system have been extensively studied 
previously by Schwartz and Rossky.\cite{BJS94d,BJS94e,BJS95a} Their
results for the  solvation response, defined as the normalized
autocorrelation of the  energy gap between the ground and excited
states, indicates a rapid 24  fs initial Gaussian component which makes
up roughly 40\% of the  response followed by a longer time 240 fs
exponential decay which  makes up most of the remainder.

As the excited state is solvated by the surrounding water molecules, 
the energy gap between the excited state and the ground state narrows 
to its equilibrium value.  Furthermore, from first order perturbation 
theory, we expect that the electronic transition rate to be inversely 
proportional to the magnitude of the energy difference between initial 
and final states.  Thus, a simple model for the excited state survival 
probability can be written
 \begin{eqnarray}
 	\frac{\partial P(t)}{\partial t} & = & -\frac{ k_{\rm eq}}{\tilde \omega(t)}P(t)
 	\label{prob-eq}
 \end{eqnarray}
subject to the initial condition $P(0)=1$.  Here, $k_{\rm eq}$ is the 
non-adiabatic transition rate for the solvated excited state and 
$\tilde \omega(t)$ is the average energy gap normalized to the average 
energy gap of the solvated excited state.  $\tilde U(t)$ is also related 
to the solvation response $S(t)$ by
\begin{eqnarray}
	\tilde \omega(t) & = &\frac{ \langle \omega_{\circ}\rangle 
	S(t) + \langle \omega_{\rm eq}\rangle (1-S(t))}{\langle \omega_{\rm eq}\rangle}.
\end{eqnarray}
The solvation response can be modeled as short time Gaussian decay
followed by a longer time exponential
relaxation,\cite{BJS94c,BJS94d,BJS94e,BJS96a}
\begin{eqnarray}
S(t) = A e^{-1/2 (t/t_G)^2} + (1-A) e^{-t/t_E}.
\end{eqnarray}
In Fig.~\ref{fig:gap}, we plot $\tilde \omega(t)$ using a parameteric fit
to the solvation response function obtained from our simulations: $A$
= 0.38, $t_G = 0.38 $ fs, $t_E = 240 $fs.  These parameters agree
exactly with those obtained by Schwartz and
Rossky\cite{BJS94c,BJS94d,BJS94e} for this system.  This comes as no
surprise since our inclusion of quantum coherence effects does not
effect the computation of the interaction forces between the excited
electron and the water molecules.   At short times, when $\tilde U(t) >
1$, the nonadiabatic transition rate will be small since the energy
gap is large.  At long times $\tilde \omega(t)\rightarrow 1$ as the energy
gap relaxes to its equilibrium value.  Thus, $k_{eq}$ is the
exponential decay constant of the excited state population once the
excited state is fully solvated.

The solution of Eq.~\ref{prob-eq} is the one parameter family of curves
given by
\begin{eqnarray}
	P(t) & = & \exp\left(-k_{\rm eq}\int_{0}^{t} \frac{ds}{\tilde 
	\omega(s)} \right)P(0).
\end{eqnarray}
Using a non-linear fitting procedure, we fit our data to this model to 
obtain an estimate of the equilibrium nonadiabatic lifetime of 241 fs 
with a $\chi^{2}= 1.707 $.  A plot of our data superimposed on this 
fit is given in Fig.~\ref{fig:swtimes}.  Interestingly enough, our data 
does not fit this simple model as nicely as the data given by Schwartz 
and Rossky.\cite{BJS94c,BJS94d,BJS94e,BJS96a}  This suggests that the
longer coherence times used in  this study imparts a non-trivial memory
dependency into the survival  probability which is inadequately captured
in Eq.~\ref{prob-eq}.  

As expected, the lifetimes reported here (Table 1) are consistently
shorter than the lifetimes reported by Schwartz and
Rossky~\cite{BJS94c,BJS94d,BJS94e,BJS96a} in which a constant 1 fs
coherence time scale throughout their simulations, thus emphasizing
the profound sensitivity of these simulations to the coherence time
scale.  Furthermore, our results are consistent with the estimated
lifetimes reported in SBPR~\cite{ERB96a} where the effective
equilibrium lifetime was estimated as a function of coherence time by
reanalyzing the transition amplitudes computed in the original
Schwartz and Rossky simulations with the coherence time scale
estimated from a frozen Gaussian approximation to the nuclear
wavefunction.

\section{Discussion}
In this paper we have briefly described the results of our work 
towards a molecular level description of quantum relaxation phenomena.  
Here we have focused exclusively upon the role that transient quantum 
mechanical coherences between the solvent and the solute play in the 
electronic relaxation of an excited solute species.  In mixed 
quantum-classical computer simulations, fundamental assumptions about 
the decay of these coherences produce direct manifestations on the 
computed quantum mechanical transition rates and must be included 
consistently in order to make realistic predictions and 
comparisons.\cite{ERB95b,ERB96a}

The computational algorithm, based upon the so-called consistent 
histories interpretation of quantum mechanics, provides both the 
molecular level underpinnings of quantum decoherence and the 
computational means for properly including decoherence effects in 
non-adiabatic quantum-molecular dynamics simulations.  According to 
the rules of ordinary quantum mechanics, a quantum system will evolve 
into a coherent superposition of alternative states.  In our 
decoherence theory, this coherence is dissipated due to the 
differences in the forces exerted on the bath by alternative states 
involved in the superposition.  Thus, on short scales, a quantum 
system in a bath will obey the rules of ordinary quantum mechanics and 
evolve into a coherent superposition of states, 
whereas on longer time scales, 
the coherences between states are diminished and the quantum system 
must be described as statistical (i.e.  incoherent) mixture of states.  
As the quantum system interacts continuously with the bath, coherences 
between states are continuously created by non-adiabatic coupling and 
damped by the divergence in the bath dynamics induced by the 
system-bath coupling.

This subtle interplay between coupling and decoherence and the 
subsequent dependency of transition rates on the decoherence time scale 
has profound implications for a variety of condensed phase chemical 
dynamics including: internal conversion and internal vibrational 
energy redistribution (in which the bath is comprised of all the modes 
of the molecules except for the one mode of interest), electronic 
energy transfer between molecules or different parts of the same 
molecule, and charge transfer reactions including proton and electron 
transfer.  In these latter examples, both the condensed phase 
environment and the internal motions of the molecules act as a bath 
which couples the quantum states together.  The decay of quantum 
coherence, which depends upon the frequencies and populations of 
the bath modes coupled to the quantum system will determine the extent 
to which the non-adiabatic coupling can act to allow the chemical 
reaction to proceed.  Changes in the spectral density due to isotopic 
changes in the bath can have a substantial impact on non-adiabatic 
chemical dynamics.\cite{ERB96a}  Furthermore, the decay of quantum 
coherence can determine the adiabaticity for a chemical reaction.

Perhaps the two major lacunae in our present theory is the explicit 
dependency upon an {\em a priori} estimate of the coherence time scale 
and the fact that this time scale remains fixed throughout the 
calculation.  As mentioned above, we estimated this time scale by 
computing the average decay time of the overlap of a product of 
Gaussian coherent states evolving on different adiabatic potential 
energy surfaces.  In this estimate, the individual widths of the 
coherent state wavefunctions centered about the initial phase space 
points of the classical nuclei are set to be proportional to the 
thermal DeBroglie wavelength of each nuclei.  Although physically 
realistic for a variety of situations, this does leave the coherence 
{\em length scale} (i.e.  the widths) as an adjustable parameter.  
While the effects of changing the coherence length scale over a 
broad range have not been systematically studied, results from our 
previous work demonstrate that the non-adiabatic transition is quite 
sensitive to changes in the coherence time scale and hence will be 
sensitive to changes in the coherence length scales.  Furthermore, as 
the bath explores various regions of the quantum potential energy 
surface, the force differences which give rise to the decay of the 
quantum coherences~\cite{ERB96a} should vary from one configuration to the next.  
Current progress is underway towards obtaining both the coherence 
length and time scale during the course of the non-adiabatic simulation 
{\em ab initio} by examining the quantum mechanical fluctuations of 
the bath particles about their stationary phase 
paths.

\acknowledgments

This work was  supported by the National Science Foundation (CHE-9504666) 
as well as the San 
Diego Supercomputing Center for computing resources. 
We also thank Professor Benjamin J. Schwartz and Oleg V. Prezhdo for 
many fruitful discussions over the course of this work.   
ERB also wishes 
to thank Professor Hans C. Andersen for his hospitality at Stanford 
University where this paper was completed.

\newpage
\begin{center}
{\bf Tables}
\end{center}
\begin{table}[h]
\caption{Excited state lifetimes for the aqueous electron. (SR= 
	Schwartz and Rossky, J. Chem. Phys. {\bf 101} 6902 (1994).  
	SBPR = Schwartz, Bittner, 
	Prezhdo, and Rossky, J. Chem. Phys. {\bf 104} 4942 (1996)). 
The coherence times used 
	in each study are listed in parentheses. }
\protect\label{Tab1}
	
\begin{tabular}{c|lll}
	\hline
	  & Present (3.1 fs ) & SR (1 fs)  & SBPR (2.8-3.1 fs)$^{({\rm a})}$ \\
	\hline
	   Median    & 338 fs & 630 & ---   \\
	  Average    & 384    & 730 & ---   \\
	 Equilibrium & 234    & 450 & 310-270   \\ 
	\hline 
\end{tabular} 

(a.) Only Equilibrium lifetimes considered. 
\end{table}
\newpage

\begin{center}
{\bf Figure Captions }
\end{center}


\begin{figure}
\caption{Energy eigenvalues and quantum populations along different 
switching paths for the electron in water.  Shown in the upper graph are 
the energy eigenvales of the occupied electronic state for two 
possible switching paths with the occupations shown in the lower 
plot.  Following promotion to the excited state, the electron makes 
a series of stochastic hopping attempts between the initial and 
final states with two possible outcomes shown.}
\protect\label{fig:2paths}
 \end{figure}
 
\begin{figure}[h]
 	\caption{Excited state cumulative survival probability as a 
 	function of time following initial excitation.  Superimposed 
 	points (A) are the switching times from the simulations.  Curve B is 
 	the fit of the data to the energy gap relaxation model given by Eq. 
 	22 and curve C is a fit of the data to a Gaussian.  See text for 
 	details.}
	\protect\label{fig:swtimes}
\end{figure}

\begin{figure}[h]
\caption{Plot of the normalized energy gap, $\omega$, between the ground and 
occupied excited state following initial excitation.}
\protect\label{fig:gap}
\end{figure}

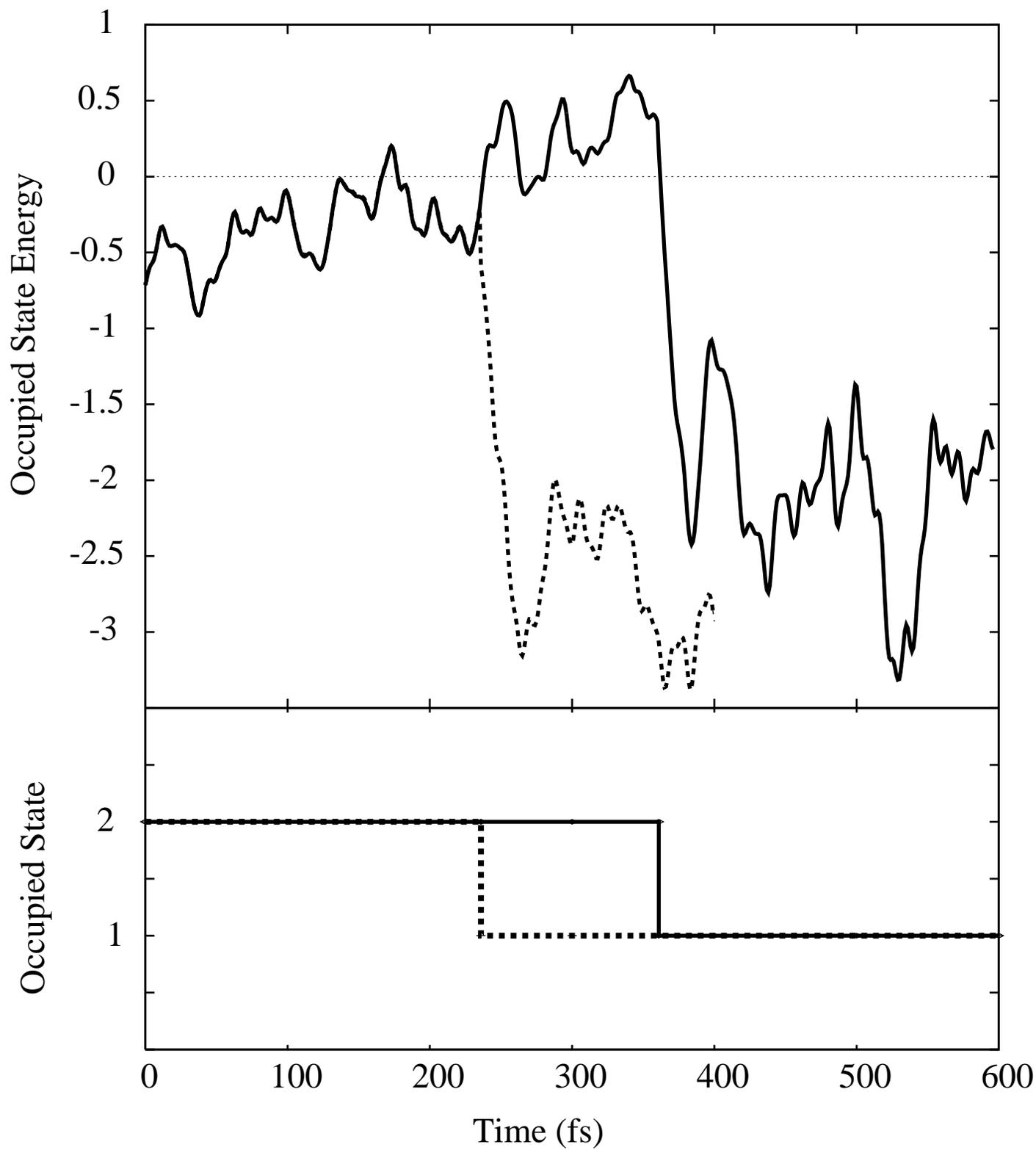
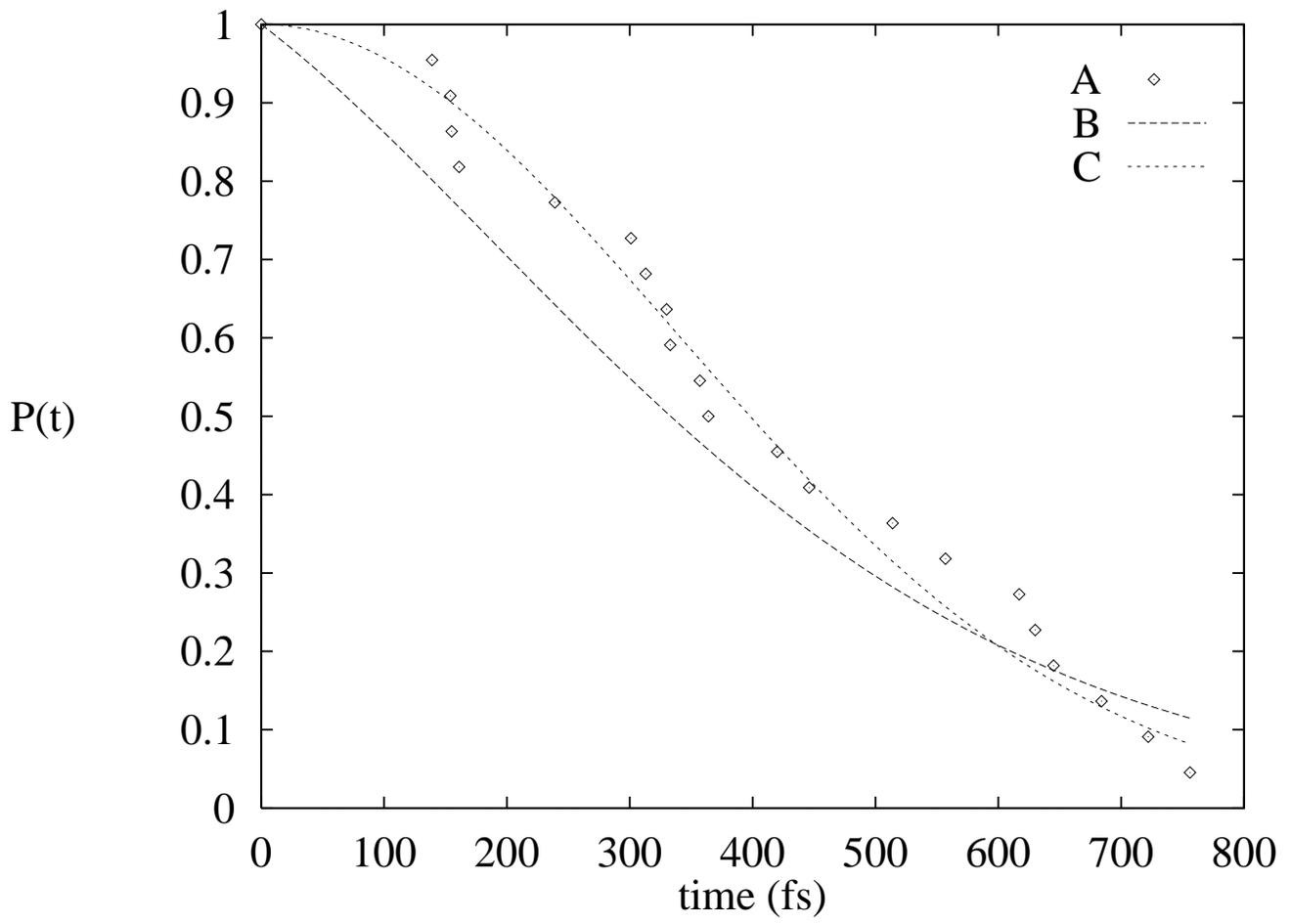
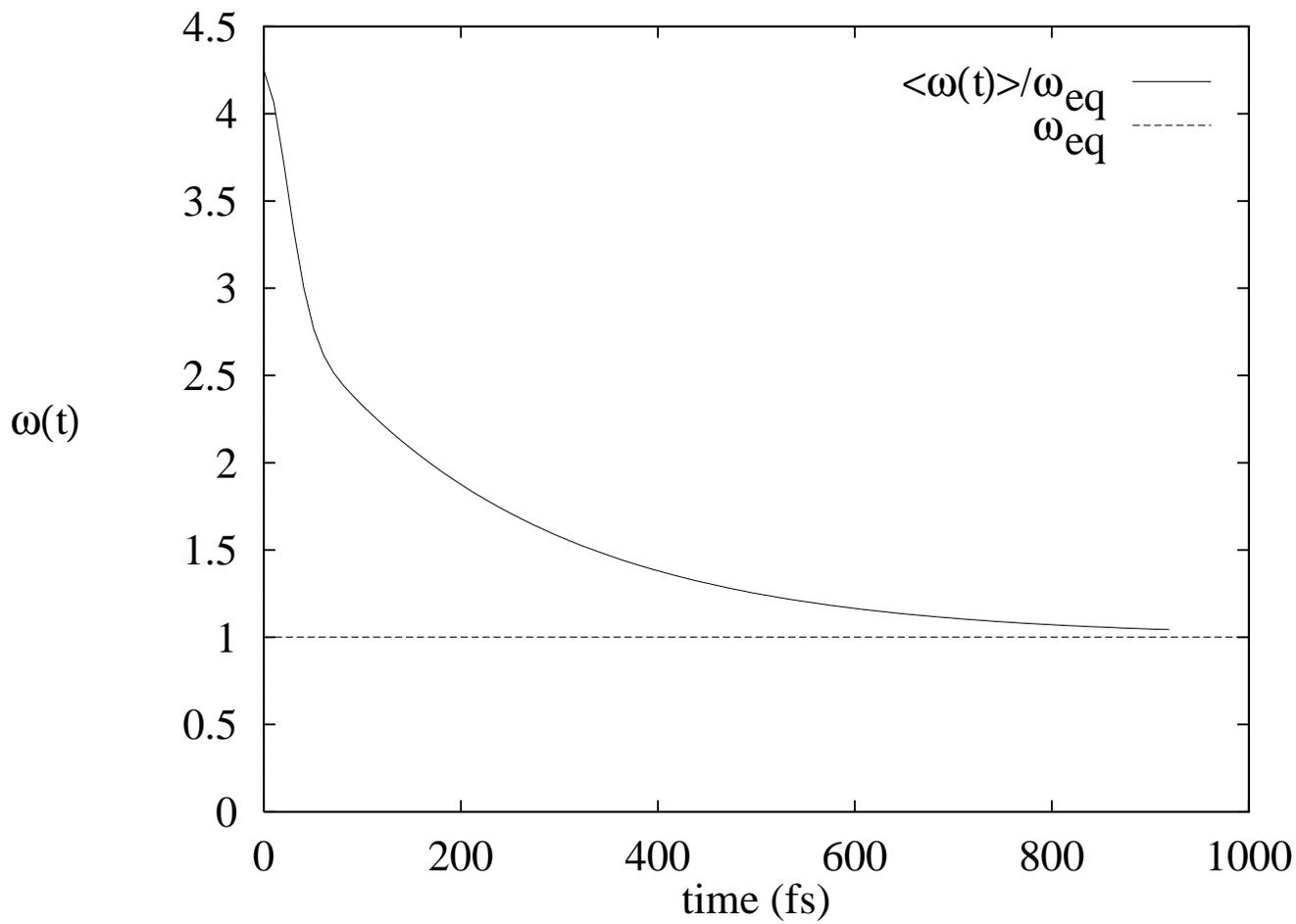
\end{document}